\documentclass[notitlepage,nofootinbib]{revtex4-1}

\usepackage{amsmath}
\usepackage{latexsym}
\usepackage{amssymb}
\usepackage{bm}
\usepackage{graphicx}
\usepackage{epstopdf}
\usepackage{comment}
\usepackage{lmodern}
\usepackage{verbatim}

\usepackage{enumerate}
\usepackage{color}
\usepackage{lineno}
\usepackage{soul}
\usepackage{xcolor}

\begin{document}

\title{Self-starting nonlinear mode-locking in random lasers}
\author{Fabrizio Antenucci$^{1,2}$, Giovanni Lerario$^{3}$, Blanca Silva Fernand\'ez$^3$, \\ Milena De Giorgi$^3$, Dario Ballarini$^3$,
Daniele Sanvitto$^{3*}$}
\author{ Luca Leuzzi$^{1,4*}$}
\address{$\phantom{.}^1$ Institut de physique théorique, Université Paris Saclay, CNRS, CEA, F-91191 Gif-sur-Yvette, France}
\address{$\phantom{.}^2$CNR-NANOTEC, Institute of Nanotechnology, Soft and Living Matter Lab., Piazzale Aldo Moro 5, I-00185,
Rome, Italy }
\address{$\phantom{.}^3$CNR-NANOTEC, Institute of Nanotechnology, Via Monteroni, I-73100 Lecce,
Italy.}
\address{$\phantom{.}^4$ Dipartimento di Fisica, Università \textrm{Sapienza}, Piazzale Aldo
Moro 5, I-00185, Rome, Italy }
\email{daniele.sanvitto@cnr.it, luca.leuzzi@cnr.it}


\begin{abstract}
\begin{center}
\par\end{center}
In ultra-fast multi-mode lasers, mode-locking is implemented by means of ad hoc devices, like saturable absorbers or modulators, allowing for very short pulses. This comes about because of nonlinear interactions induced among modes at different, well equispaced, frequencies. Theory predicts that the same locking of modes would occur in random lasers but, in absence of any device, its detection is unfeasible so far. Because of the general interest in the phenomenology and understanding of random lasers and, moreover, because it is a first example of self-starting mode-locking we devise and test a way to measure such peculiar non-linear coupling. 
Through a detailed  analysis of multi-mode correlations we provide clear evidence for the occurrence of nonlinear mode-coupling in the cavity-less random laser made of a powder of GaAs crystals and its self-starting mode-locking nature. 
The behavior of multi-point correlations among intensity peaks is tested against the nonlinear frequency matching condition equivalent to the one underlying phase-locking in ordered ultrafast lasers. 
Non-trivially large multi-point correlations are clearly observed for spatially overlapping resonances and turn out to  sensitively depend on the frequency matching being satisfied, eventually demonstrating the occurrence of non-linear mode-locked mode-coupling. 
\end{abstract}
\maketitle

When light propagates through a random medium, scattering reduces
information about whatever lies across the medium, fog and clouds
being everyday-life examples. The electromagnetic field is composed
by many interfering wave modes, providing a complicated emission pattern
as light undergoes multiple scattering. In the so-called random lasers
\cite{Cao99,Cao03r,Wiersma08,Andreasen11,Cao98,Cao99,Cao03r,Anni04,Cao00,vanderMolen06,vanderMolen07,El-Dardiry10,Tulek10,Augustine15}, 
this random scattering is used to reach the population inversion activating the lasing action.
The random laser device is made of an optically active
medium and randomly placed light scatterers. The medium provides the
gain for population inversion under external pumping. The scatterers 
provide the high refraction index and the feedback mechanism of multiple
scattering, playing a role analogous to cavity mirrors in standard
lasers, and  leading to amplification by stimulated emission.
The same material can both sustain the gain and the scattering \cite{Cao98,Cao99,Cao03r,Anni04},
else two apart components with complementary functionality can be
combined\cite{Cao00,vanderMolen06,vanderMolen07,El-Dardiry10,Tulek10,Augustine15,Viola18}.

Standard multi-mode laser theory has shown that that the dominant mode interaction
above threshold is highly non-linear\cite{Haus00}, nonlinearity
being represented by multi-mode couplings and characterized by mode-locking.
In the random laser case, couplings are predicted to be disordered,
both from the point of view of the interaction network and for what
concerns the coupling values. Therefore, 
cross-mode interactions understanding is a very debated topic and fundamental questions still need to be answered: how strong are the mode
couplings? What is their sign?  How many modes are simultaneously involved
in each interaction?

Clearly, modes must spatially overlap to manifest
mode locking 
\cite{Antenucci15a,Antenucci15b,Antenucci16}. 
This
has been observed in experiments on specifically designed random lasers, where pairwise (therefore, linear) interaction
manifests as the consequence of a two modes competition for sharing
their mutual mode intensities within the same optical volume 
\cite{Leonetti11}.
Spatial overlap is not, however, a sufficient condition for interaction,
nor it provides any information about the coupling values. At the
same time, the exact structure of the spatial distribution of the
modes in commonly used random lasers is hard to be determined, which
makes a quantitative analysis of the interacting parameters hard to
be obtained.
Therefore, we have developed a theoretical analysis - making use of statistical mechanics - of random systems of interacting light modes providing information about the mode-coupling constants. 

Thanks to this statistical physics analysis on the emission spectra of a GaAs powder-based random laser, we not only experimentally demonstrate the non-linear coupling of spatially overlapping modes, but we also provide evidence of its mode-locking nature.

\section*{Results}
\subsection*{\em Mode-locking in ordered and random lasers} 
Despite the mode-locking phenomenon is known to be nonlinear and light
modes are expected to be coupled, the mechanism and nature of this
nonlinear coupling in random lasers has never been experimentally
tested. On the other hand, the theory for stationary regimes in an
active random medium under external pumping leads to a description
in terms of an effective stochastic non-linear potential dynamics
for the mode slow amplitudes $a(t)$ (more information in Sec. A of Supplementary information)
whose Hamiltonian reads 
\begin{eqnarray}
{\mathcal{H}} & = & -\sum_{\mathbf{k}_{2}\mid\text{FMC}(\mathbf{k})}g_{k_{1}k_{2}}^{(2)}\,a_{k_{1}}a_{k_{2}}^{*}-\frac{1}{2}\sum_{\mathbf{k}_{4}\mid\text{FMC}(\mathbf{k})}g_{k_{1}k_{2}k_{3}k_{4}}^{(4)}\,a_{k_{1}}a_{k_{2}}^{*}a_{k_{3}}a_{k_{4}}^{*}+\mbox{c. c.}\label{eq:H}
\end{eqnarray}
where the acronym FMC on sums stays for the nonlinear Frequency Matching
Condition
\begin{eqnarray}
|\omega_{k_1}-\omega_{k_2}+\omega_{k_3}-\omega_{k_4}|<\gamma \quad; \qquad \gamma \equiv \sum_{j=1}^4 \gamma_{k_j}
\label{eq:fmc}
\end{eqnarray}
where $\gamma$'s are the linewidths of the resonances  angular frequency domain.
Besides this  requirement,
further complexity of the mode interaction is hidden inside the $g$
coupling coefficients, 
\begin{eqnarray}
g_{k_{1}k_{2}k_{3}k_{4}}^{(4)}\propto\int_{V}d\bm{r}~\hat{\chi}^{(3)}(\bm{r};\omega_{k_{1}},\omega_{k_{2}},\omega_{k_{3}},\omega_{k_{4}})\cdot\bm{E}_{k_{1}}(\bm{r})\bm{E}_{k_{2}}(\bm{r})\bm{E}_{k_{3}}(\bm{r})\bm{E}_{k_{4}}(\bm{r})\label{eq:coup}
\label{def:randcoup}
\end{eqnarray}
where $\hat{\chi}^{(3)}$ is the nonlinear susceptibility tensor of
the medium and $\bm{E}_{k}(\bm{r})$ the slow amplitude mode of frequency
$\omega_{k}$.

In standard lasers, when the system is pumped above threshold, the
FMC induces phase-locking  \cite{Marruzzo15,Antenucci15c}.
This is responsible for the onset of ultra-short pulses in standard
multimode lasers \cite{Haus00,Gordon02,Gat04}, in which the resonating
cavities are designed in such a way that mode frequencies of the gain
medium have a comb-like distribution \cite{Udem02,Baltuska03,Schliesser06}.
To reach mode-locking, nonlinear devices are employed, such as saturable
absorbers, for passive mode-locking, or modulators
synchronized with the resonator round trip, for active mode-locking\cite{Haus00}.
In random lasers no evidence has been obtained so far about the occurrence
of mode-locking in connection with mode-coupling. Indeed in the random
case no \textit{ad hoc} device is present in the resonator and even
the definition of \textit{resonator} is far from straightforward\cite{Lethokov68}.
Consequently, {\it mode-locking would be}, in case, {\it a self-starting
phenomenon} due to the randomness of scatterers' position and the optical
etherogeneity of the random medium. 

In principle, the direct way to
identify a possible mode-locking in random lasers would be to look
for a temporal pulse, composed by modes at different
frequencies, with a non-trivially locked phase. However, such a putative
pulse might realistically be longer than typical pulses in standard
mode-locking lasers and modes at different frequencies would contribute
differently and in an uncontrolled way \cite{Soukoulis02}. { {Moreover,
pulses do not form unless mode frequencies are regularly separated and
mode couplings $g^{(4)}$ take (mostly) positive values.}}
Indeed, it can be theoretically proved  that even when FMC, cf. Eq. (\ref{eq:fmc}), is satisfied,
{{continuously distributed frequencies \cite{Marruzzo15} or 
a non-negligible fraction  of the random mode interactions given in Eq. \eqref{def:randcoup}  \cite{Gradenigo19},
 \footnote{Negative optical response is supposed to occur  in the so-called \textit{glassy} random lasers \cite{Antenucci15a,Antenucci15f,Antenucci16,Marruzzo16} }
prevent the onset of
laser pulses.}}
\typeout{do not even allow for phase-locking, i.e., $\phi(\omega)\sim\omega$ does not follow FMC of Eq. (\ref{eq:fmc}). }

Since the direct observation
of mode-locking via optical pulses is neither practical
nor conclusive, in this work we demonstrate a different approach to
detect mode-locking in random lasers, based on multi-point cross-correlation
measurements. 

Using this method we prove 
  that in a random laser modes at different wavelenghts interact nonlinearly. Moreover, their interaction is mode-locked, i. e.,  their frequencies satisfy the  matching
condition reported in Eq. (\ref{eq:fmc}).

\subsection*{\em Data analysis} 
\label{sec:res} 
Our random laser is composed by a thin deposition of GaAs powder, cf. Methods. 
A Gaussian laser beam (780 nm excitation wavelength) illuminates the sample propagating perpendicular to the deposition plane (x,y). 
 The detection line is along the $z$ direction in transmission configuration (i.e., at the opposite side of the sample with respect to the excitation line).
The sample thickness is irregular in the $z$
direction, though always thinner than   $100$ $\mu$m. 

The emission intensity in random lasers
is typically far too low to allow a good resolution in all three dimensions
($x,y,\omega$) within a single shot. In our experiments we have 
 emission spectra from a given slice of the sample ($10$
$\mu$m wide) resolved in the $x$ coordinate ($x,\omega$) for $100$, $1000$ and $10000$ shots, corresponding
to $10$, $100$, $1000$ ms integration time. This acquisition
times allowed us to obtain a high enough spectral resolution to adequately
probe  the presence of nonlinear interaction and the role played
by the FMC on the random laser emission by means of statistical information criteria.

Our analysis of the experimental spectra consists in: (i) identifying
all resonances in all acquired spectra at all available positions
$x$; (ii) selecting strongly correlated resonance sets out of the distribution
of correlations among all possible mode sets; (iii) verifying that strongly
correlated sets satisfy FMC.

\subsubsection*{Resonances identification }

\label{sec:4a} 
\begin{figure}[t!]
\centering 
\includegraphics[width=0.54\linewidth]{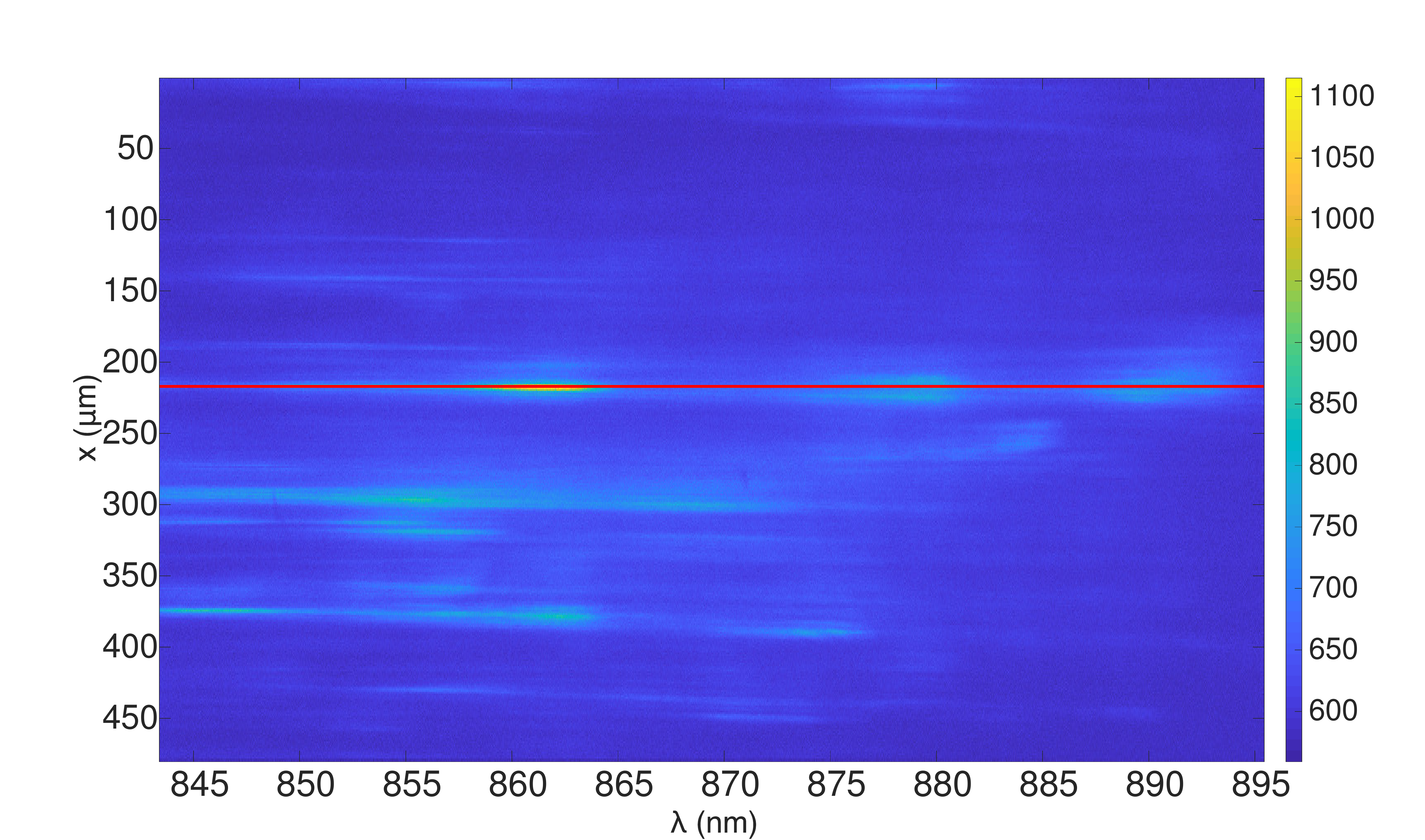}
\includegraphics[width=0.44\linewidth]{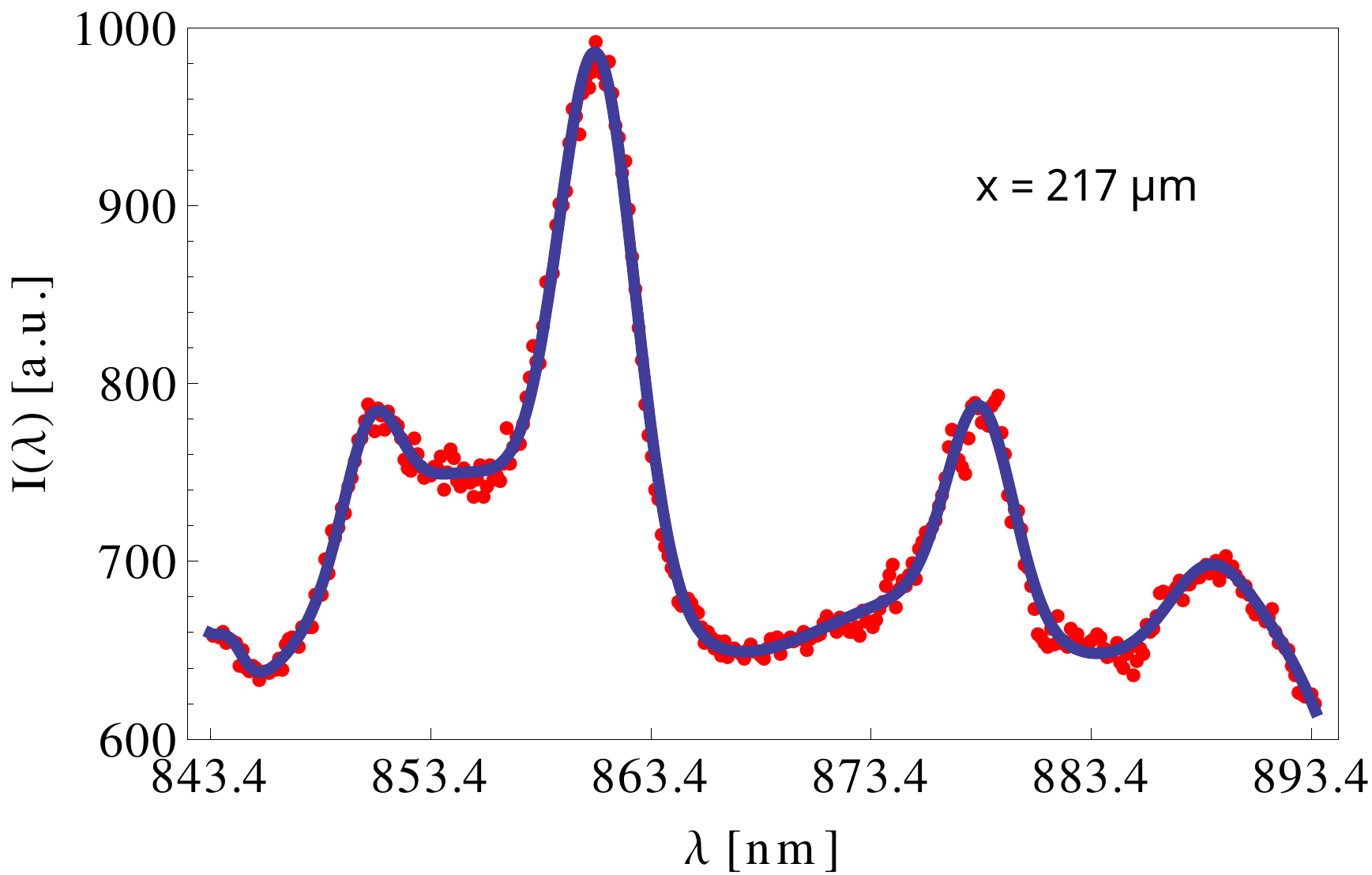}
\caption{(a) Instance of an intensity spectrum vs position and wavelength.
(b) Instance of multi-Gaussian interpolation of the intensity spectrum
of GaAs crystal powder in a single $10$ms data acquisition time ($100$ shots) at
coordinate $x=217~\mu$m, corresponding to the red section in (a). The minimal number $N_{G}$ of Gaussian curves
used is determined by the interpolating set yielding the least Akaike
parameter. Depending on the $x$ coordinate, $N_{G}$ turns out to
vary between $5$ and $10$ {[}see Supplementary information{]}. }
\label{fig:mGaussian_fit} 
\end{figure}

Peaks in the spectra are identified by performing
multiple fitting with linear combinations of a variable number of
Gaussians (details are reported in Supplementary information).
To avoid overfitting, the optimal set of curves is chosen according
to the Akaike Information Criterion \cite{Akaike74}. An instance
of the outcome of our fitting procedure is reported in Fig. \ref{fig:mGaussian_fit},
right panel, where we plot raw data compared to multi-Gaussian interpolating
functions. Eventually, we build a complete list of all resonances
for each spectrum produced in each one of the different data acquisitions
$t=1,\ldots,N_{{\rm spectra}}=1000$ in the series of measurements.
Each intensity peak $k$ of the spectrum $t$ is determined by its
frequency $\omega_{k}$, its linewidth $\gamma_{k}$, its position
$x_{k}$ and the FWHM $\Delta x_{k}$ in its position coordinate:
$I_{k}^{(t)}\equiv I^{(t)}(x_{k},\Delta x_{k};\omega_{k},\gamma_{k})$.

\subsubsection*{Strong correlation discrimination}
\label{sec:4b}
Of each set of intensity peaks we compute the normalized fourth order
cumulants of their intensities $c_{4}(\omega_{j},\omega_{k},\omega_{l},\omega_{m})$,
cf. Methods. Since there are many spurious
effects that may contribute to a correlation among modes, in order
to identify anomalously large correlations we need a reference for
the background correlation. %
\begin{figure}[t!]
\includegraphics[height=0.22\linewidth]{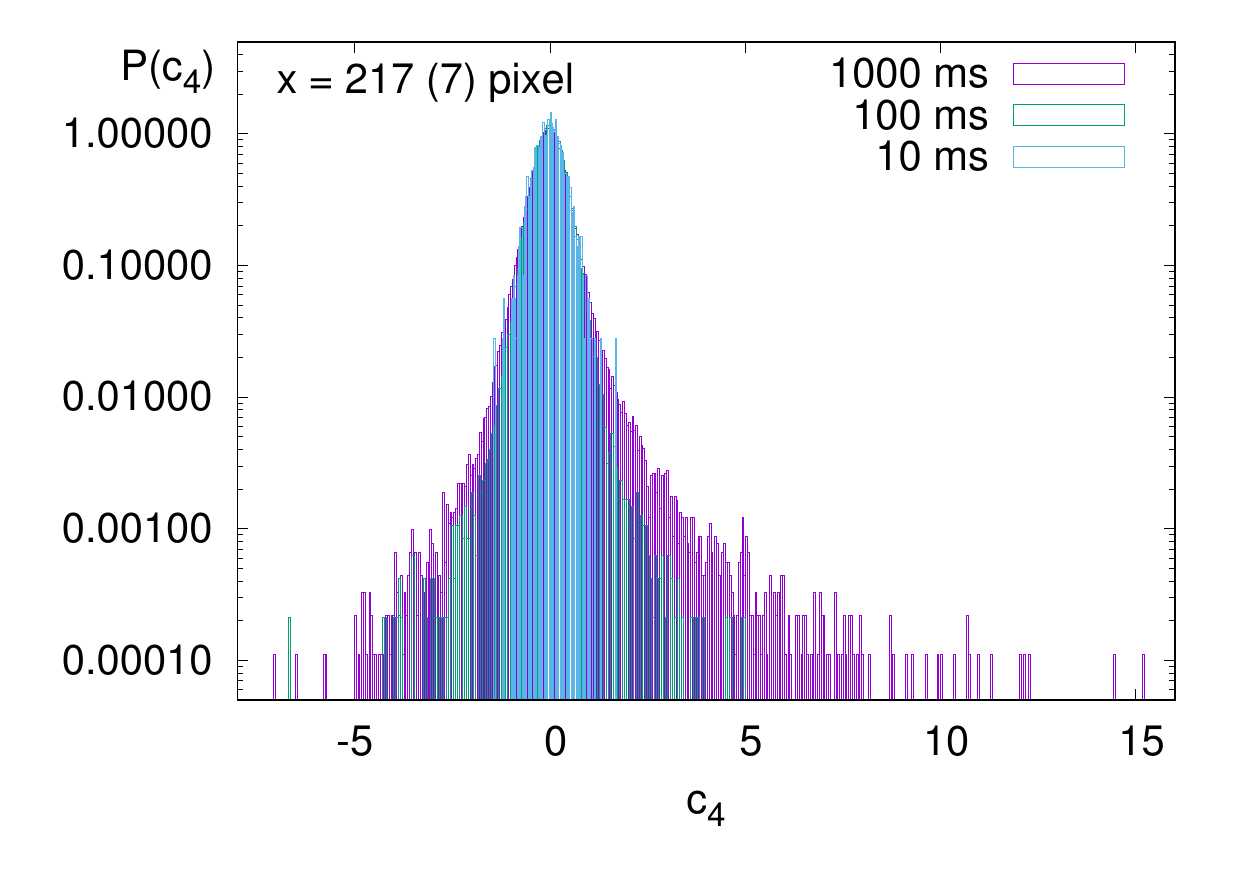}
\includegraphics[height=0.22\linewidth]{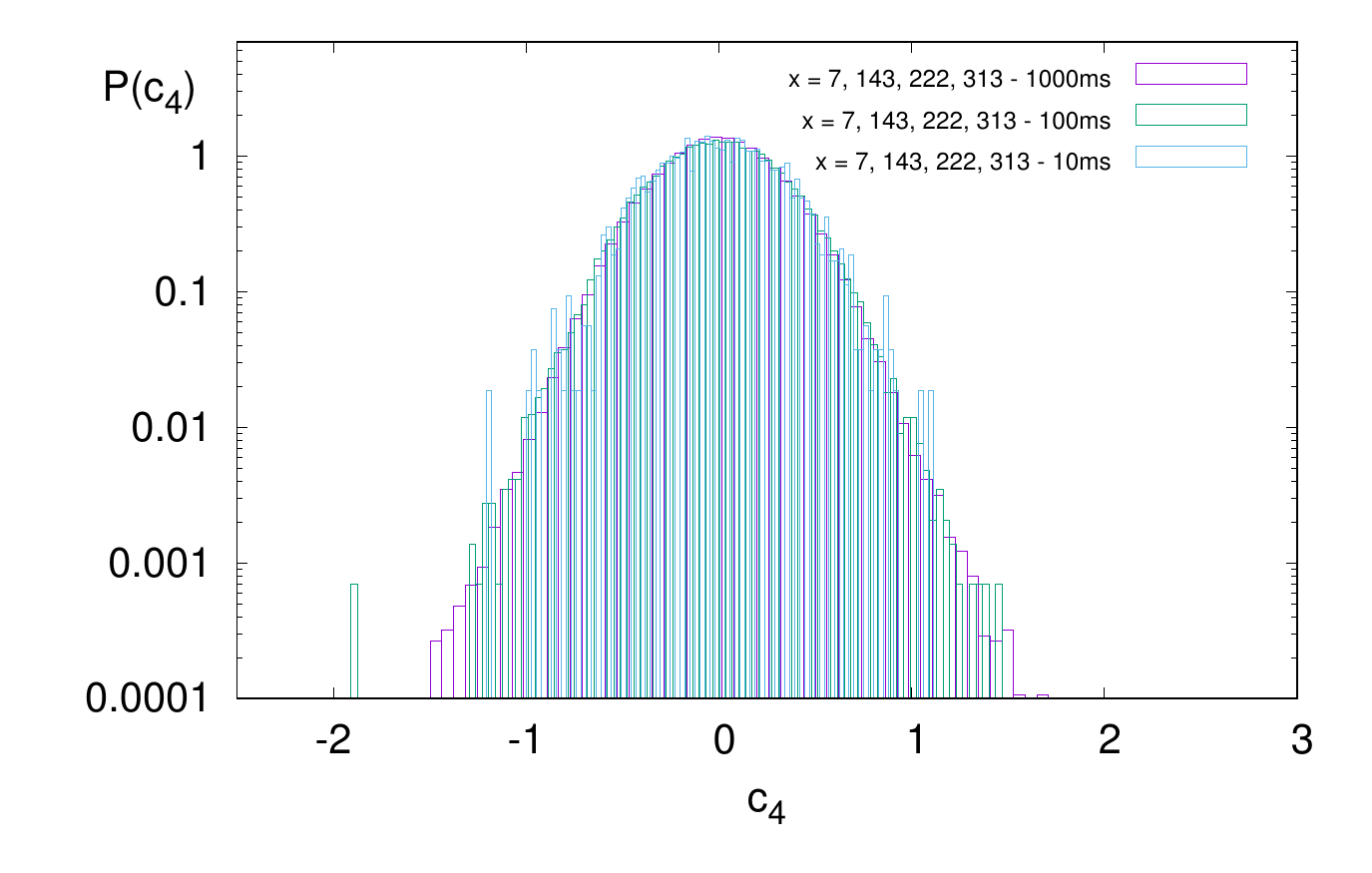}
\includegraphics[height=0.22\linewidth]{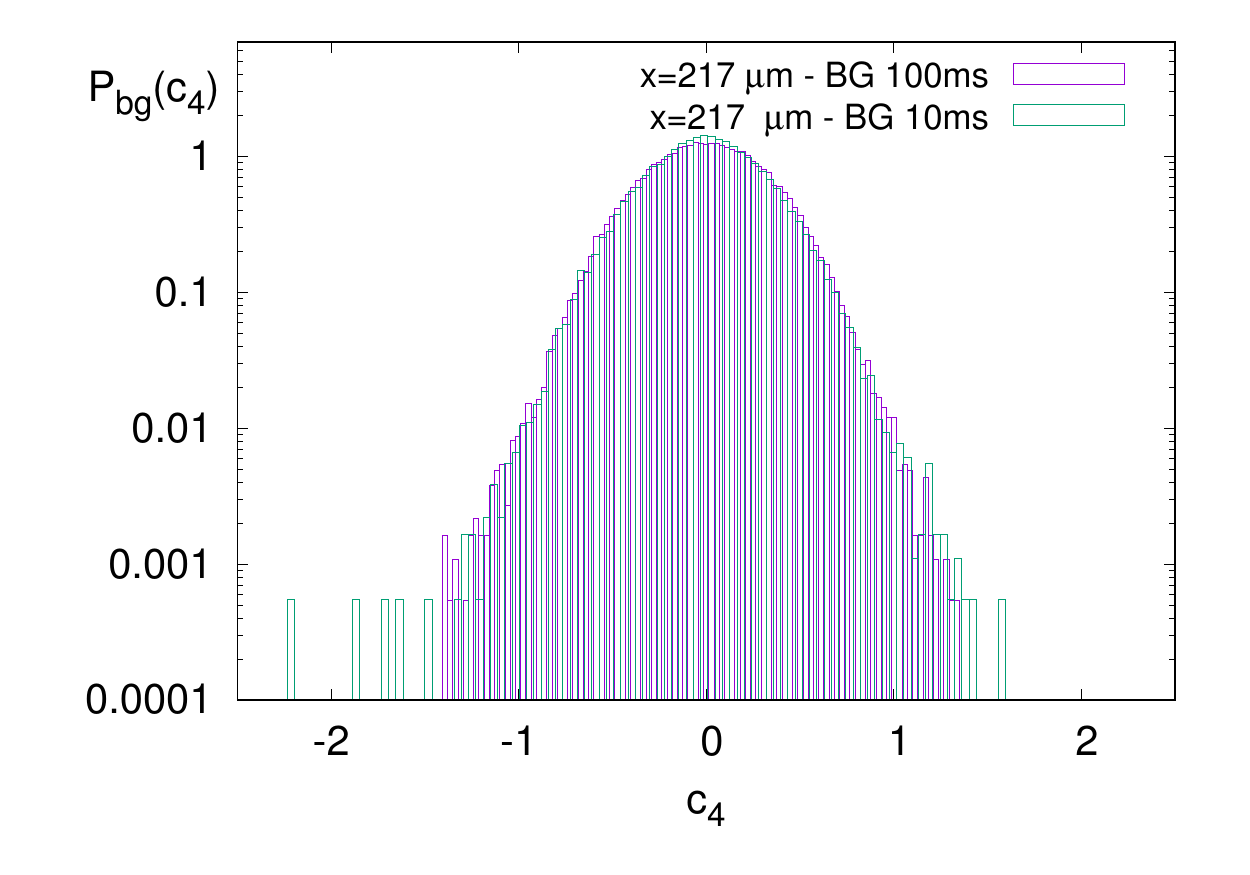}
\caption{Normalized distribution of $c_{4}$ for acquisition times $10$, $100$,
$1000$ ms. (a) $c_{4}$ taken at the same shot, same $x=210~\mu$m. (b) $c_{4}$ taken at
the same shot, but different positions  $x$. (c) $c_{4}$ taken at the different
shots, same $x$.}
\label{fig:c4_distribs} 
\end{figure}
One can observe that the largest correlation values (in the tails of the distributions)  turn out to be
attained always on those sets whose resonances taken at the same time overlap in the $x$
position (SOIR), with respect to NOIR and background correlations.
Moreover, comparing the distributions in Fig. \ref{fig:c4_distribs}
as the number of acquired emissions increases, the dominion of possible
values for the SOIR correlations extends its extremes in the low probability
tails. This is not the case, on the contrary, for NOIR and background
$c_{4}$, insensitive to the change in acquisition time. 
We, thus, compute three kinds of multi-point correlations: 
\begin{itemize}
\item {SOIR}, the correlations of all quadruplets composed by possibly
spatially self overlapping intensity resonances, i.e., occurring at
the same planar position $x$; \footnote{We recall that the spectra  results from the integration over $10$ $\mu$m along the $y$ direction and a deposition thickness lower than $100$ $\mu$m  so that not all resonances
at same $x$ are guaranteed to be actually spatially overlapping,
see Supplementary information.}
\item {NOIR}, the correlations quadruplets composed by non overlapping
intensity resonances, i.e., occurring at well distinct $x$ positions
in the same spectral data acquisition; 
\item {BG}, the background correlations among resonances pertaining to
different spectra, i.e., acquired after different pump shots. 
\end{itemize}
In  Fig. \ref{fig:c4_distribs} (a) we display the probability
distributions of the SOIR four-point correlation
functions $c_{4}$ for $100$, $1000$
and $10000$ shots. {{Sets of SOIR are candidate to be nonlinearly interacting. }}
In the center panel of Fig. \ref{fig:c4_distribs} we
display the NOIR $c_{4}$. The latter set might still be composed
by interacting modes in an extended mode scenario \cite{Fallert09}. Finally, in the
right panel of Fig. \ref{fig:c4_distribs} the BG $c_{4}$ are plotted.

\begin{figure}[t!]
\includegraphics[width=0.49\linewidth]{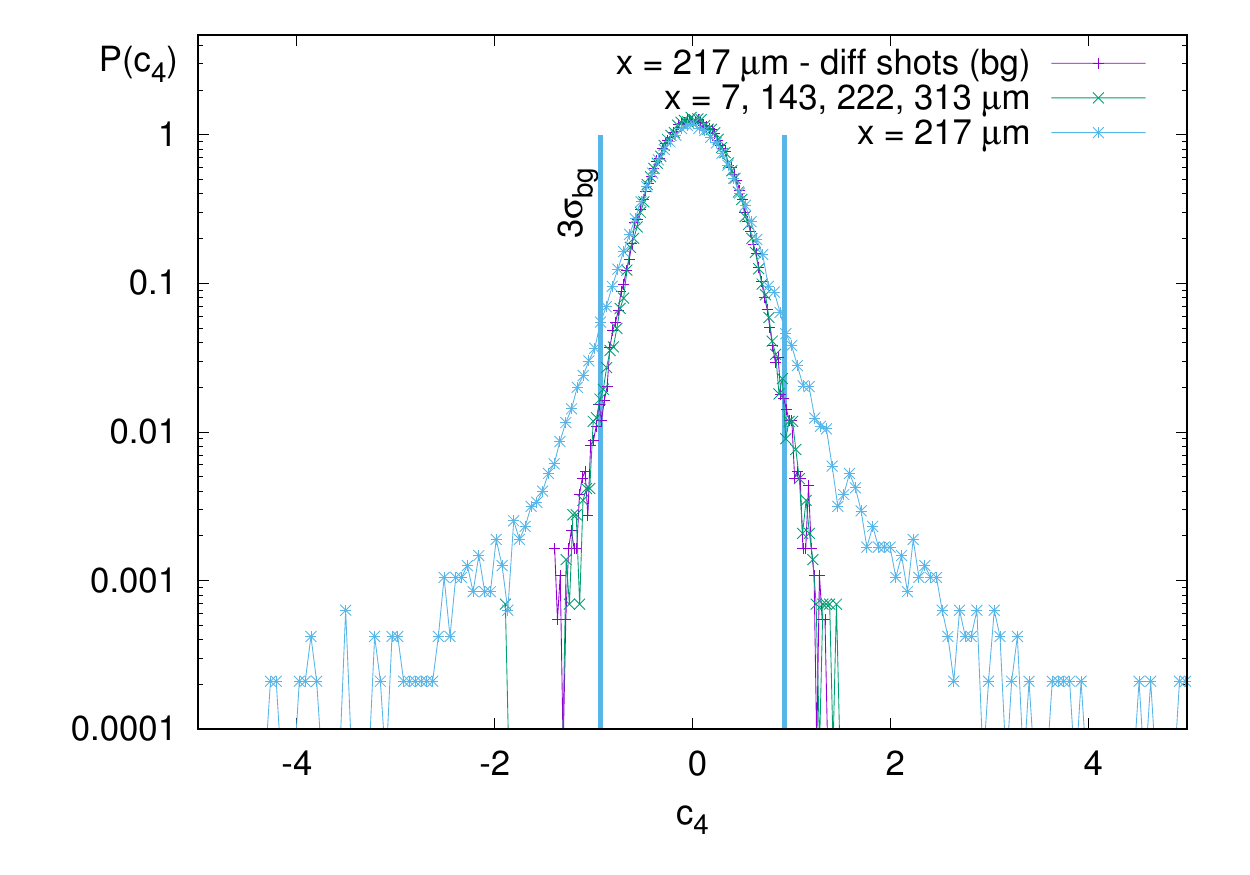}
\caption{Normalized distributions of $c_{4}$ at acquisition time $100$ ms
for background correlations, non-overlapping resonance correlations
and spatial overlapping resonance correlations. The vertical lines
correspond to $\pm3\sigma$ of the background correlation distribution.}
\label{fig:c_distribs_3s} 
\end{figure}

With high confidence, we, finally, operatively identify nonlinearly
\textit{interacting} sets of modes as those whose multi-mode correlation
is larger - in absolute value - than the $3\sigma$ of the background
correlation distribution. In Fig. \ref{fig:c_distribs_3s} we superimpose
instances of the normalized distributions for the background, the
NOIR and the SOIR correlations for an acquisition time of $100$ ms, clearly
showing that the tails of the SOIR extend well beyond the $3\,\sigma$
of the other two.

The presence of these interacting sets of modes is our first result:
 spatially overlapping modes  interact nonlinearly
in the random lasing regime.

\subsubsection*{Frequency matching role  in strongly correlated modes}
\label{sec:4c}
\begin{figure}[t!]
\includegraphics[width=0.49\linewidth]{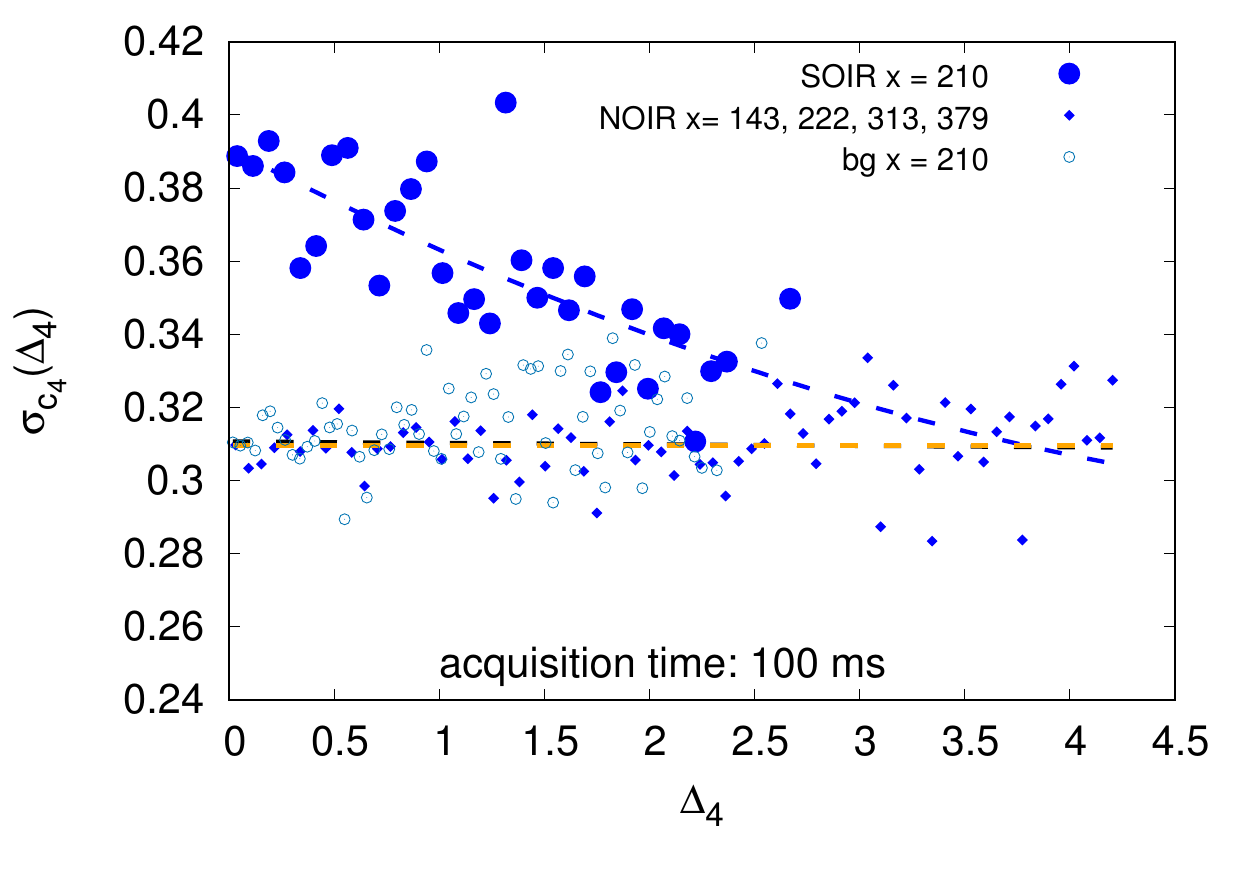}
\includegraphics[width=0.49\linewidth]{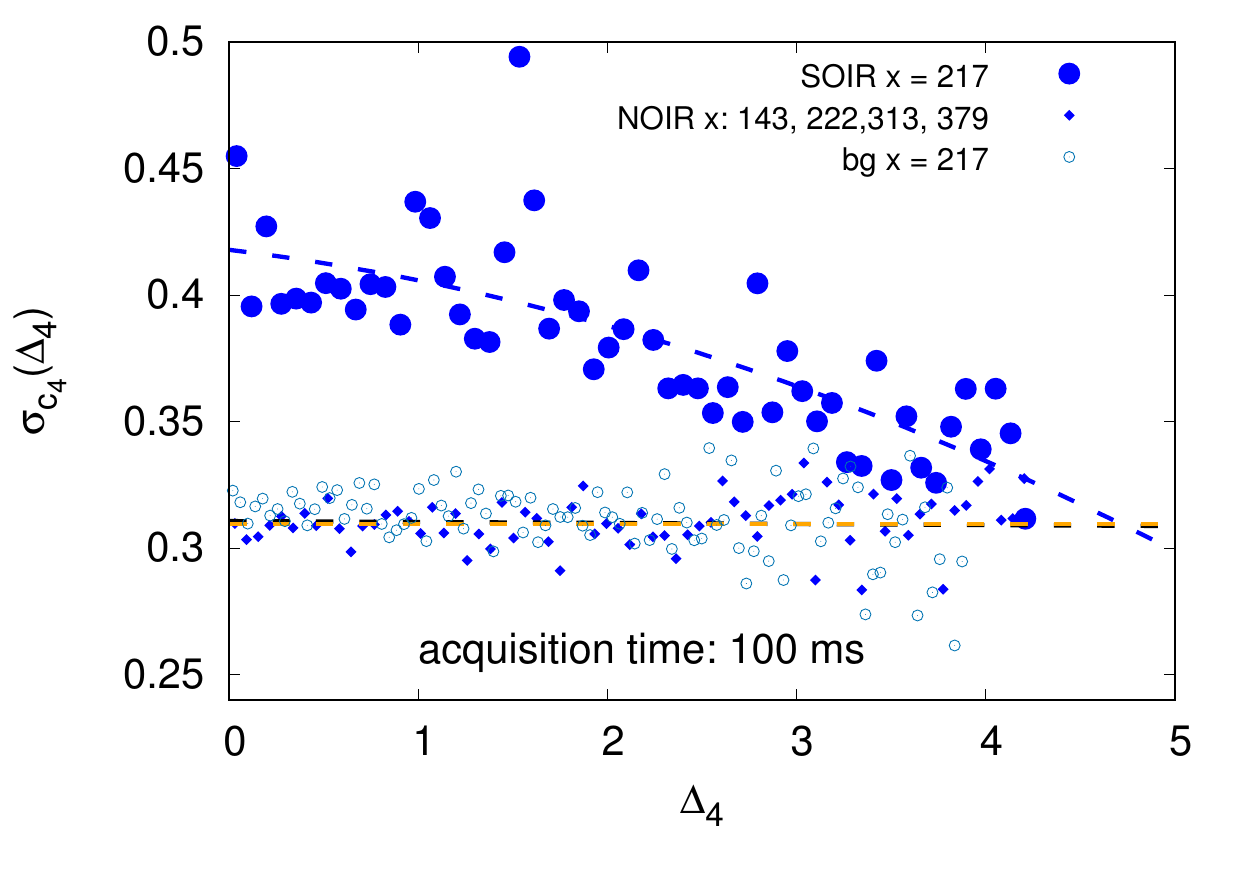}
\caption{Mean square displacement of the four-point correlation $c_{4}$ values
vs. $\Delta_{4}$ at $100$ ms acquisition time for background correlation
(bg), non-overlapping resonances and space-overlapping resonances.
Left: at pixel $x=210(7)~\mu$m. Right: $x=217(7)~\mu$m. }
\label{fig:c4_vs_fmc} 
\end{figure}

Eventually, we are interested in testing whether the frequencies of
interacting modes satisfy FMC. This
is indeed, a signature for self-starting mode-locking in the GaAs
random laser. To make FMC  quantitative, cf. Eq. (\ref{eq:fmc}), we introduce a
``FMC parameter'' against which we can straightforward test multi-mode
correlations. In the case of the $4$-mode correlation, taking for illustrative
purpose modes $1,2,3$ and $4$, this is \footnote{There are actually three non-equivalent permutations of indices in
Eq. (\ref{eq:fmc}). We always consider all three distinct permutations
and take the smallest $\Delta_{4}$.}
\begin{eqnarray}
\Delta_{4}\equiv\frac{|\omega_{1}-\omega_{2}+\omega_{3}-\omega_{4}|}{\gamma_{1}+\gamma_{2}+\gamma_{3}+\gamma_{4}}\label{def:Delta}
\end{eqnarray}
In Fig. \ref{fig:c4_vs_fmc} the mean square displacement $\sigma_{c_{4}}$
of the correlations among modes within a given $\Delta_{4}$ interval
are displayed. For correlations computed from a series of $1000$
spectra, each one acquired in $100$ ms, we plot $\sigma_{c_{4}}$ the distributions 
of the SOIR, of the NOIR and of the background correlations.
It can be observed that \textit{no} dependence on $\Delta_{4}$ is
shown for BG and NOIR correlations. On the contrary, the $\sigma_{c_{4}}$'s
of SOIR correlation distributions depend on $\Delta_{4}$. In particular,
non-trivially strong correlations only occur among the SOIR, and only for small $\Delta_{4}$,
signaling that in those sets of modes - those most probably coupled
- their interaction strength depends on how well FMC is satisfied.

\begin{figure}[t!]
\centering 
 \includegraphics[width=0.95\linewidth]{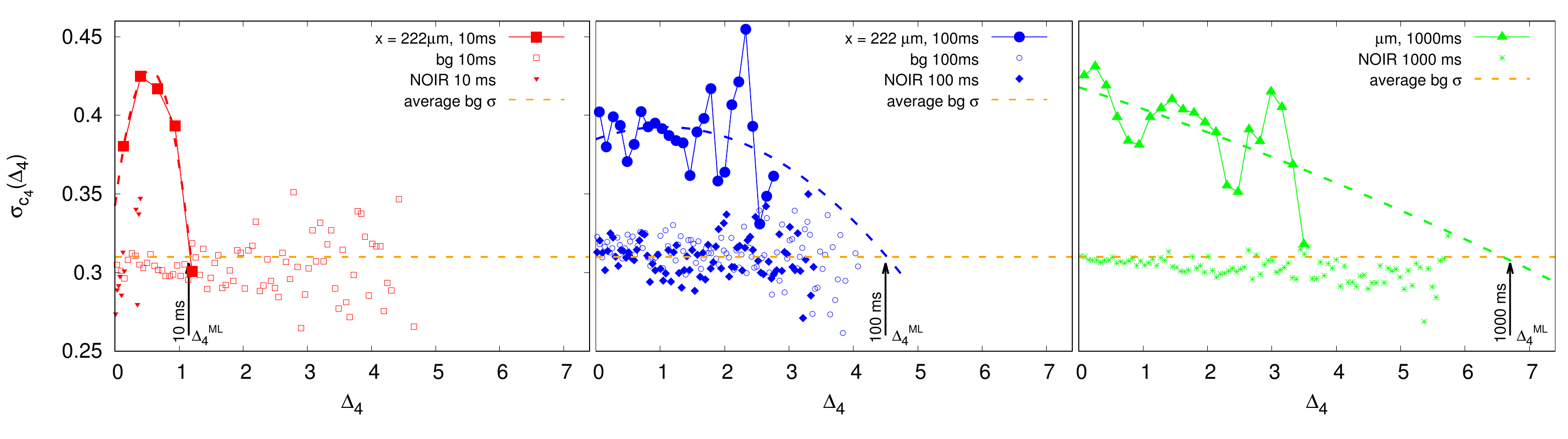}
\caption{Mean square displacements $\sigma_{c_{4}}$ of the distributions of
four-resonance correlation values $c_{4}$ at fixed $\Delta_{4}$
intervals.  For SOIR quadruplets at $x=222~\mu$m
we use large full points,
for instances of background (non-iteracting) quadruplets we use small
empty points and for the instances of NOIR quadruplets small full
points. The $\sigma_{c_{4}}$'s are plotted versus the FMC parameter
$\Delta_{4}$ for acquisition times $10$, $100$ and $1000$ ms, corresponding
to a spectral integration over, respectively, $100$ (red), $1000$ (blue) and $10000$ (green)
pumping shots. Dashed lines are parabolic interpolations of
SOIR $\sigma(\Delta)$ behaviors. Background and NOIR correlations
do not show any dependence on FMC, whereas SOIR correlation distributions,
with large tails at $\Delta_{4}\simeq 0$ corresponding to interacting
sets, tend to shrink as the FMC is progressively relaxed. {{The boundary value
 $\Delta_{4}^{{\rm ML}}$ at which SOIR $\sigma$'s decreases to values
of the order of background $\sigma$'s depends on the acquisition
time. In particular, 
$\Delta_{4}^{{\rm ML}}$ decreases with the number of shots, towards
the expected limit of $\Delta_{4}^{{\rm ML}}\lesssim 1$ for a single shot. }}}
\label{fig:c4_interaction_spot} 
\end{figure}

This behavior occurs at all acquisition times used in experiments,
as it can be observed in Fig. \ref{fig:c4_interaction_spot}. 
In particular,
 one can observe that, decreasing the acquisition time, i.e., decreasing
the number of recorded photon emissions after each pumping laser shot,
the threshold value of $\Delta^{\rm ML}_{4}$ below which surely interacting
modes can be neatly discriminated from background correlation decreases.

 Ideally, to analyse mode-locking,
for every single shot one would like to have the intensity of emission
measured as a function of space $(x,y)$ in the sample plane, to control
the spatial overlap, and contemporarily resolved in angular frequency
$\omega$, to check the FMC, Eq. (\ref{eq:fmc}). 

Although the integration over many shots (the emission intensity is too low to allow the resolution of the energy-space emission map of a single shot experiment), we can clearly observe the shrinking of $\Delta^{\rm ML}_4$ with decreasing number of shots, consistently with the upper theoretical limit of $\Delta^{\rm th}_4=1$ for single shot experiments.

The evidence that only SOIR show $\Delta$-dependent correlations
is a clear indication of the onset of nonlinear mode-locking in random
lasers and our main result: sets of modes satisfying FMC are much
strongly correlated than sets of modes not satisfying FMC. The former
non-trivially large correlations is to be accounted for by the interaction of the
modes. The latter cannot be distinguished from $\Delta$-independent
background correlations.

\section*{Discussion} 
In the present work we considered multi-mode correlations among spatially
overlapping intensity resonances (SOIR), spatially non-overlapping
intensity resonances (NOIR) and background correlations among resonances
in independent spectra (BG).
Firstly, by compared analysis of sets of background
correlations and correlations among NOIR we cannot appreciate any
difference in the behavior of their distributions, as reported
in Figs. \ref{fig:c4_distribs} and \ref{fig:c_distribs_3s}.
Since correlations among NOIR are not any larger than background correlations
we cannot discriminate possible long-range non-linear mode-coupling
with respect to noise. 

This is not the case for SOIR multi-point correlations.
Calibrating $P_{{\rm SOIR}}(c_{4})$ by means of the background correlation
distributions, we  identify  interacting quadruplets
as those composed by modes whose intensity fluctuations correlation
lies in the tails of their distribution, actually extending beyond
$3~\sigma_{{\rm BG}}$ of the Gaussian interpolation of the $c_{4}$
BG distribution, as shown in Fig. \ref{fig:c_distribs_3s}.

The same behavior is found also if we change the pumping power, as
far as the random laser is above threshold and clear resonances can
be distinguished in the space-energy spectrum (Supplementary information).

As observed in Section Results, cf. Figs. \ref{fig:c4_vs_fmc}, \ref{fig:c4_interaction_spot},
the frequency matching condition, Eq. (\ref{eq:fmc}), appears to play
a determinant role in the distribution of the $c_{4}$ values of 
interacting set of modes. Moreover, we observe that the shorter the
acquisition time, the smaller the range of values of $\Delta_{4}$
at which large correlation occurs. 
According to Eq. (\ref{eq:fmc}),
interaction between modes would be allowed only to modes whose energies
satisfy the FMC relationship. In terms of data reported in Figs. \ref{fig:c4_interaction_spot}
this would imply that interacting mode sets should appear for $\Delta\lesssim O(1)$.
Given the statistical nature of (i) the modes identification and (ii)
the interaction recognition from anomalously large multi-point correlations,
the outcome is strongly compatible with such a requirement. 

This observation
is a strong evidence in favour of the occurrence of mode-locking in
random lasers, that is, the same mechanisms behind the nonlinear mode
coupling in standard, ordered, multimode lasers,  though without any \textit{ad hoc}
device like a saturable absorber or a modulator. It is a self-starting
mechanism induced by randomness.

As a last remark we recall that in the ordered case mode-locking is
responsible for ultra-fast pulses. On the contrary in random lasers, no train of pulses
is present,  because the  distribution
of frequencies is random, rather than comb-like \cite{Udem02}, preventing the
rise of a pulse even in presence of frequency matching and phase-locking.
Indeed, the Fourier transform in time does not produce a modulated
signal with a short envelope \cite{Marruzzo15}. Actually, also
in presence of almost equispaced resonances it would be
extremely difficult to identify a pulse shorter than the pumping laser
pulse and, in the subclass of optically random media displaying \textit{glassy} random
lasers, this might not be feasible at all \cite{Gradenigo19}. To unveil
the self-starting mechanism beyond the just demonstrated locking of
modes in random lasers mandatorily requires the identification of
mode phases. We believe that the presented  results might be a significant step to stimulate and lead the  theoretical understanding and the experimental procedures necessary to  provide a protocol to determine mode phases in random lasers.


\section*{Materials and Methods}
\subsection*{Samples}

\label{methods:sample}

The sample is made by grinding a piece of GaAs wafer - bathed in methanol
- in a pestle and mortar, in order to obtain a paste with grain typically
smaller than $10~\mu$m. The resulting paste is then highly diluted
in methanol and deposited by drop casting on a glass substrate. During
the deposition, the glass substrate is placed on a heater to have
a fast evaporation of the solvent. The packing density and sample
thickness is increased by repeating $50$ times the drop casting process.
Obviously, the sample is highly inhomogeneous. However, the thickness
never exceeds $100~\mu$m. A second glass slide covers the sample,
which is finally sealed with parafilm on the sides.

\subsection*{Experimental setup and measurements}

\label{methods:exp}

The laser source is a $30$ fs pulsed laser at $780$ nm with repetition
rate $10$ KHz. The excitation line is orthogonal to the sample surface,
while the detection line is along the opposite side of the sample
(i. e., transmission configuration). Adjusting the spot size and excitation
power, we can roughly control the number of active random lasers loops.
The Gaussian spot size is tuned to about $300~\mu$m (FWHM); this
spot size guarantees a total number of emitters that are easily distinguishable
when the real space emission map is projected on the CCD camera. The
detection line consists of two plano convex lenses projecting the
sample plane at the spectrometer entrance (so in focus on the CCD
camera). The spectrometer slits aperture allows the selection of a
vertical slice of the emission space map and the energy resolution
of the emission spectrum. The experiments have been performed with
a slit aperture corresponding to $10~\mu$m horizontal selection
on the sample surface, in order to have the selection of a single
emitter along the horizontal axis. The overall numerical aperture
of the detection line is $0.64$.

\subsection*{Multi-point correlation definition}

\label{app:c4def} The fourth order connected correlation function  of intensity peaks $I_j(\omega_j)$ reads:
\begin{eqnarray}
C_{4}(\omega_{j},\omega_{k},\omega_{l},\omega_{m}) & = & \langle I_{j}I_{k}I_{l}I_{m}\rangle-\langle I_{j}I_{k}I_{l}\rangle\langle I_{m}\rangle-\langle I_{j}I_{k}I_{m}\rangle\langle I_{l}\rangle-\langle I_{j}I_{m}I_{l}\rangle\langle I_{k}\rangle-\langle I_{m}I_{k}I_{l}\rangle\langle I_{j}\rangle
\nonumber
\\
 &  & -\langle I_{j}I_{k}\rangle\langle I_{l}I_{m}\rangle-\langle I_{j}I_{l}\rangle\langle I_{k}I_{m}\rangle-\langle I_{j}I_{m}\rangle\langle I_{k}I_{l}\rangle+2\langle I_{j}I_{k}\rangle\langle I_{l}\rangle\langle I_{m}\rangle+2\langle I_{j}I_{l}\rangle\langle I_{k}\rangle\langle I_{m}\rangle\nonumber \\
 &  & +2\langle I_{j}I_{m}\rangle\langle I_{k}\rangle\langle I_{l}\rangle+2\langle I_{k}I_{l}\rangle\langle I_{j}\rangle\langle I_{m}\rangle+2\langle I_{k}I_{m}\rangle\langle I_{j}\rangle\langle I_{l}\rangle+2\langle I_{l}I_{m}\rangle\langle I_{j}\rangle\langle I_{k}\rangle-6\langle I_{j}\rangle\langle I_{k}\rangle\langle I_{l}\rangle\langle I_{m}\rangle\nonumber 
\end{eqnarray}
where the average is taken over  the statistical sample of all the combinations of the same set of modes displayed in experiments at fixed external condition and stable pumping. 
We, further, normalize $C_{4}$ 
to the mean square displacements of the intensities of the modes,
i.e., 
\begin{eqnarray}
\nonumber
c_{4}(\omega_{j},\omega_{k},\omega_{l},\omega_{m}) & \equiv & \frac{C_{4}(\omega_{j},\omega_{k},\omega_{l},\omega_{m})}{\sigma_{j}(\omega_{j})\sigma_{k}(\omega_{k})\sigma_{l}(\omega_{l})\sigma_{m}(\omega_{m})}\\
\sigma_{j}(\omega_{j}) & = & \sqrt{\langle\left(I_{j}-\langle I_{j}\rangle\right)^{2}\rangle}
\nonumber
\end{eqnarray}


\section*{Supplementary materials}

\subsection{Theoretical modeling}

\label{app:model}

The dynamics of the electromagnetic field is suitably expressed in
the slow mode decomposition 
\begin{eqnarray}
\bm{E}(\bm{r},t)=\sum_{k}a_{k}(t)e^{\imath\omega t}\bm{E}_{k}(\bm{r})+\mbox{c.c.}
\end{eqnarray}
where the modes $\bm{E}_{k}(\bm{r})$ of angular frequency $\omega_{k}$
are such that their complex amplitudes $a(t)$ evolve slowly with respect to $\omega^{-1}$
and follow a nonlinear stochastic dynamics. In standard mode-locked
lasers with passive mode-locking induced by a saturable absorber, e. g.,
the phasors dynamics is given by the so-called Haus master equation
\cite{Haus00}, that, in the angular frequency domain \cite{Gordon02},
reads: 
\begin{eqnarray}
\dot{a}_{k_{1}}(t)=(g_{k_{1}}-\ell_{k_{1}}+\imath D_{k_{1}})a_{k_{1}}(t)+(\gamma+i\delta)\sum_{k_{2}k_{3}k_{4}}^{{\rm FMC}}a_{k_{2}}(t)a_{k_{3}}^{*}(t)a_{k_{4}}(t)+\eta_{k_{1}}(t)
\label{eq:master}
\end{eqnarray}
where $g_{k}$, $\ell_{k}$ and $D_{k}$ are, respectively, the frequency-dependent
components of the gain, the loss and the dispersion velocity. In the
case of passive mode-locking, the real part of the non-linear coefficient $\gamma$
represents the self-amplitude modulation coefficient of the saturable
absorber and the imaginary part $\delta$ represents the  coefficient of
the self-phase-modulation caused by the Kerr-effect. The acronym FMC
on sums stays for Frequency Matching Condition, cf. Tab. \ref{tab:fmc}
and Eq. (2)$_{\text{MT}}$ in the main text, arising in the approach
leading to the master equation (\ref{eq:master}), after averaging
out single mode's fast oscillations $\sim e^{\imath\omega_{n}t}$
\cite{Haus00,Gordon02,Gat04,Katz06,Antenucci15c,Antenucci15d,Antenucci15b}.
The white noise $\eta_{n}(t)$ is a stochastic variable representing
the contribution of spontaneous emission, linked to the thermal kinetic
energy of the atoms through its covariance 
\begin{equation}
\langle\eta_{k}(t)\eta_{n}(t')\rangle\propto T\delta_{kn}\delta(t-t')
\end{equation}
where $T$ is the temperature. For arbitrary spatial distribution
of modes $\bm{E}(\bm{r})$ and heterogeneous susceptibility, i.e.,
when the system is intrinsically random, starting from quantum dynamical
Jaynes-Cumming equations, downgrading from quantum creation-annihilation
operators to classical complex-valued amplitudes, taking into account  spontaneous
emission and  gain saturation, a generalized phasor dynamic
equation is recovered for random lasers \cite{Antenucci16}: 
\begin{eqnarray}
\dot{a}_{k_{1}}(t)=\sum_{k_{2}}^{{\rm FMC}}g_{k_{1}k_{2}}^{(2)}a_{k_{2}}(t)+\sum_{k_{2}k_{3}k_{4}}^{{\rm FMC}}g_{k_{1}k_{2}k_{3}k_{4}}^{(4)}~a_{k_{2}}(t)a_{k_{3}}^{*}(t)a_{k_{4}}(t)+\eta_{k_{1}}(t)\label{eq:a_Lange}
\end{eqnarray}
where we have considered only the first nonlinear term satisfying
time reversal symmetry in the electromagnetic phasor's dynamics. Further
terms would only perturbatively modify the leading behavior of the
fourth order term. Odd terms like those related to the $\chi^{(2)}$
optical susceptibility, occurring in non-centrometric potentials,
can also be included theoretically but practically will play no role
because of the usually limited wavelength dominion in the intensity spectra of the random lasers
\footnote{No chance of second harmonic generation, for instance.}.
The complexity of the mode interaction is hidden inside the $g$ coefficients in Eq. (3)$_{\text{MT}}$ of the main text.
Finally, recognizing in Eq. (\ref{eq:a_Lange})
a potential Langevin equation, the effective phasor Hamiltonian of
Eq. (1)$_{\text{MT}}$ in the main text is derived. 


\begin{table}[h!]
\centering{}%
\begin{tabular}{|c|c||c|}
\hline 
$\#$ $\omega$'s  & indices  & FMC \tabularnewline
\hline 
$2$  & $\mathbf{k}_{2}\equiv{k_{1},k_{2}}$ & $|\omega_{k_{1}}-\omega_{k_{2}}|<\gamma_{k_{1}}+\gamma_{k_{2}}$\tabularnewline
\hline 
$3$  & $\mathbf{k}_{3}={k_{1},k_{2},k_{3}}$  & $|2\omega_{k_{1}}-\omega_{k_{2}}-\omega_{k_{3}}|<2\gamma_{k_{1}}+\gamma_{k_{2}}+\gamma_{k_{3}}$\tabularnewline
\hline 
$4$  & $\mathbf{k}_{4}={k_{1},k_{2},k_{3},k_{4}}$  & $\left.\begin{array}{c}
|\omega_{k_{1}}-\omega_{k_{2}}+\omega_{k_{3}}-\omega_{k_{4}}|\label{eq:FMC4}\\
|\omega_{k_{2}}-\omega_{k_{1}}+\omega_{k_{3}}-\omega_{k_{4}}|\\
|\omega_{k_{1}}-\omega_{k_{3}}+\omega_{k_{2}}-\omega_{k_{4}}|
\end{array}\right\} <\gamma_{k_{1}}+\gamma_{k_{2}}+\gamma_{k_{3}}+\gamma_{k_{4}}$\tabularnewline
\hline 
\end{tabular}\caption{Summary of all possible FMC's for modes interacting via Eq. (1).
Pairwise, three- and four-body terms are reported. }
\label{tab:fmc} 
\end{table}


\subsection{Observable definitions and data analysis}

\subsubsection*{Identification of the intensity peaks of the modes by multi Gaussian
interpolation}

\label{methods:GaussFit} We herewith describe the fitting procedure
composing the resonances identification step in Section Results in the main text.
In a first series
of measurements, because of a  large refinement in energy with respect to the behavior
of emission spectra, we have first coarse-grained the position-energy
grid binning four pixels in the $\lambda$ direction. This corresponds to a 
resolution of $0.15$ nm in the wavelength (and $1.2\mu$m in position).
We have $335$ pixels in the wavelength
direction, with the lowest extreme being $\lambda=843.4116$ nm and
the total spectral width amounting to $52.0075$ nm. Considering an
approximately constant spacing $\Delta\lambda$, this means that 
\begin{eqnarray}
\lambda[\mbox{nm}]=52.0075/334*(\mbox{pixel}-1)+843.4116\quad;\qquad\mbox{pixel}\in[1:335]
\end{eqnarray}

In the second series of experiments the wavelength pixels are only
$672$, with resolution already of $0.149$ nm (and $0.9$ $\mu$m
in position) and there was no need for binning.

We, then performed multi-Gaussian interpolations of spectra $I(x;\lambda)$
at fixed position $x=1,\ldots,N_{x}$. We denote the number of Gaussians
employed by $N_{G}$ and we estimate two parameters for each Gaussian
$n=1,\ldots,N_{G}$, mean $\bar{\lambda}_{n}$ and variance $\sigma_{n}^{2}$,
as well as a relative weight $w_{n}$, for a total number of parameters
$K=3N_{G}$. At each position $x$, and for each $N_{G}$, we compute
the log-likelihood 
\begin{equation}
\ln\mathcal{L}\left(\lambda|\{w_{n}\},\{\bar{\lambda}_{n}\},\{\sigma_{n}\}\right)=\ln\sum_{n=1}^{N_{G}}w_{n}\exp\left[-\frac{\left(\lambda-\bar{\lambda}_{n}\right)^{2}}{2\sigma_{n}^{2}}\right]
\end{equation}
The best fit at each $x$ is the one whose parameter estimators for $\{w_n,\bar \lambda_n, \sigma_n\}$ maximize the log-likelihood $\ln{\cal L}$
 or, equivalently, minimize the so-called the Akaike parameter 
\[
A=2K-2\ln{\cal L}
\]
built with the least number of Gaussians. i.e., the least
$K$. This is the Akaike Information Criterion to avoid overfitting and underfitting.

To each experiment $t$ of the series, a grid of mode intensities
is associated: $I^{(t)}(x,\Delta x;\lambda;\Delta\lambda)$, where
$\Delta x$ and $\Delta\lambda$ are the Full Width Half Maximum of
the interpolated distributions around, respectively, $x$ and $\lambda$ ($\Delta\lambda_n = 2\sqrt{2\ln 2}\sigma_n$).

We consider the same single mode $n$ as present in the spectra of
two different experiments $t_{A}$ and $t_{B}$ of the series if 
\begin{eqnarray*}
 &  & x_{n}^{\{t_{A}\}}=x_{n}^{\{t_{B}\}}\\
 &  & |\lambda_{n}^{\{t_{A}\}}-\lambda_{n}^{\{t_{B}\}}|<\delta\lambda\\
 &  & 2\frac{\Delta\lambda_{n}^{A}-\Delta\lambda_{n}^{B}}{\Delta\lambda_{n}^{A}+\Delta\lambda_{n}^{B}}<\bar{\delta}\lambda
\end{eqnarray*}
where the absolute and relative uncertainties $\delta \lambda$ and $\bar\delta \lambda$ are chosen depending on the resolution required.
$\Delta x$ is not considered in the analysis but it is of the order
of 6 pixels. I. e., 7 ${\mu}$m in the first series of experiments
and $4$ $\mu$m in the second one.

In  both series of experiments we used  $\delta\lambda=1.5$ nm and $\bar{\delta}\lambda=0.1$.
For spectra of lower resolution and less intense peaks we also considered
a rougher coarse graining, where two parameters in the mode identification
are $\delta\lambda=4.5$ nm and $\bar{\delta}\lambda=0.3$.
We resorted to this less precise approximation exclusively when the
modes' statistics is too low to provide clear behaviors of the distributions
of $c_{4}$ values. This is the case, e.g., in some of the data with
acquisition time of $10$ ms and low pumping energy in the second
series of experiments (see Sec. \ref{sec:P}).

\subsubsection*{{Localization of resonances in the monochromator vertical direction
$y$}}

Having nearby $x$ coordinates for the modes is a {\em necessary}
condition to have spatial overlap of the intensity peaks. Indeed,
mode's extensions should also overlap in the $y$ direction, the horizontal
direction of the slit, and in the $z$ direction, the thickness of
the sample. Knowing the total range in $y$ and $z$, we can estimate
the probability that if four modes have the same $x$ they will be
effectively mutually overlapping also in $y$ and $z$. Let us call
$L_{y}$ the total $y$-range, equal to the slit opening ($L_y\simeq 10~\mu$m)
and  $\Delta \equiv \Delta y\simeq\Delta x  = 7~\mu$m the  typical extension of the most peaked resonances that we analyzed.
Let us, then, take as indicative average depth of the sample $L_{z}\simeq50~\mu$m, the
total $z$-range. The probability that four modes of roughly the
same extension $\Delta$ at the same $x$ share a non-zero spatial
intersection along the $y$ (or $z$) coordinate, in a total range
$L_{y}$ (or $L_{z}$) (enough larger than $\Delta$), can be estimated as
\begin{eqnarray}
p_{4}^{(y,z)}(\Delta_{w},L_{w}) & = & 8\frac{\Delta_{w}^{3}}{L_{w}^{3}};\qquad\qquad w=y,z
\end{eqnarray}
This is independent from the coordinates of the modes. For the setup
and the sample we used the probability of having a four mode overlap
at given $x$ position is, thus, 
\begin{equation}
p_{4}^{(x)}=p_{4}^{(y)}(7,10)\times p_{4}(7,50)\simeq 0.060
\end{equation}
This means that once we know that four resonances at different wavelength
of the acquired spectra are at the same point $x$, they will be actually
overlapping in 3D space only a fraction $p_{4}^{(x)}$ of the times.
In the results shown in this paper, we always have at least $\sim100$
quadruplets of resonant modes at given $x$ coordinate, so that we
can expect with high probability that in all cases there are at least
few spatially overlapping sets for all the cases shown. At the same
time, the fact that the probability of overlapping is relatively low
($\lesssim 10\%$) explains the presence of many low-interacting sets even in the SOIR setup and 
makes our analysis of the background correlations essential
to extract the signal coming from the few spatially overlapping sets.

\subsection{Multi-set mode correlation vs. statistics}

\label{app:stat_corr}
\begin{figure}[h!]
\centering 
 \includegraphics[width=.99\textwidth]{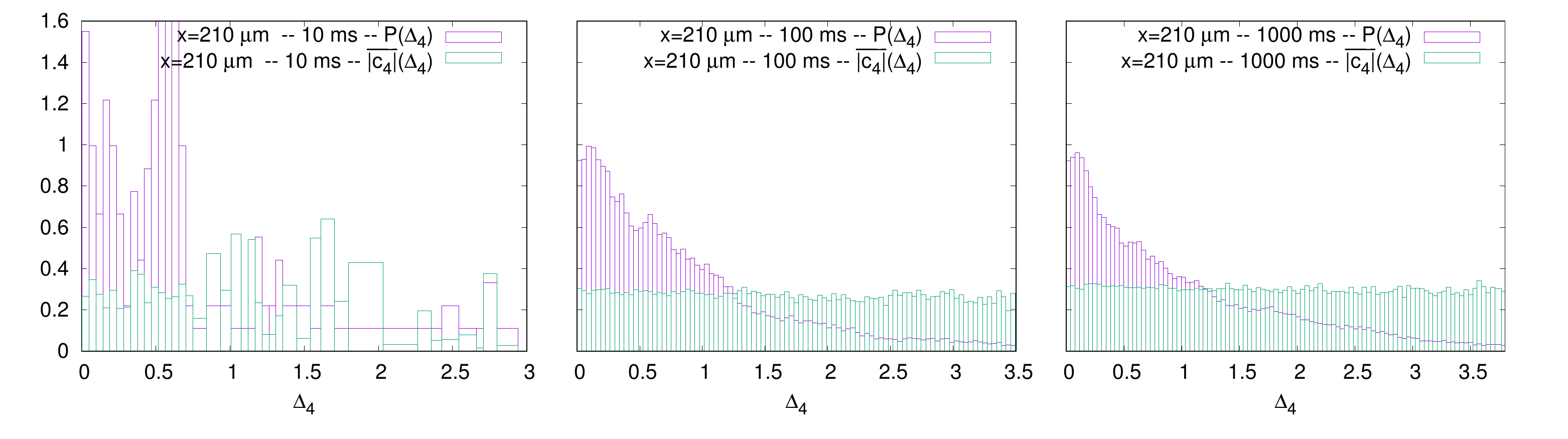}
 \caption{
Normalized histogram $P_{q}(\Delta_{4})$ of the number
of SOIR quadruplets (at position $x=210~\mu$m) whose frequencies yield
a given $\Delta_{4}$ compared to a average value of the modulus of
the correlation $c_{4}$. }
\label{fig:Histo_N_C1} 
\end{figure}

We further considered how finite sample size may affect the results shown in paper.
In particular, note that typically the number of possible quadruplets decreases rapidly as a function of $\Delta_{4}$.
We therefore performed a further test to verify whether a high (low) correlation
$c_{4}$ at a given $\Delta_{4}$ value might be related to
the presence of a large (small) number $N_{q}$ of sets of modes for
the same frequency combination $\Delta_{4}$. In Fig.~\ref{fig:Histo_N_C1}
the rescaled histograms of $N_{q}$ as function of $\Delta_{4}$ are
plotted together  with the average value of the modulus
of the correlation $|c_4|$ at the same $\Delta_{4}$: no special correlation can
be observed.

\begin{figure}[h!]
\centering 
 \includegraphics[width=.49\textwidth]{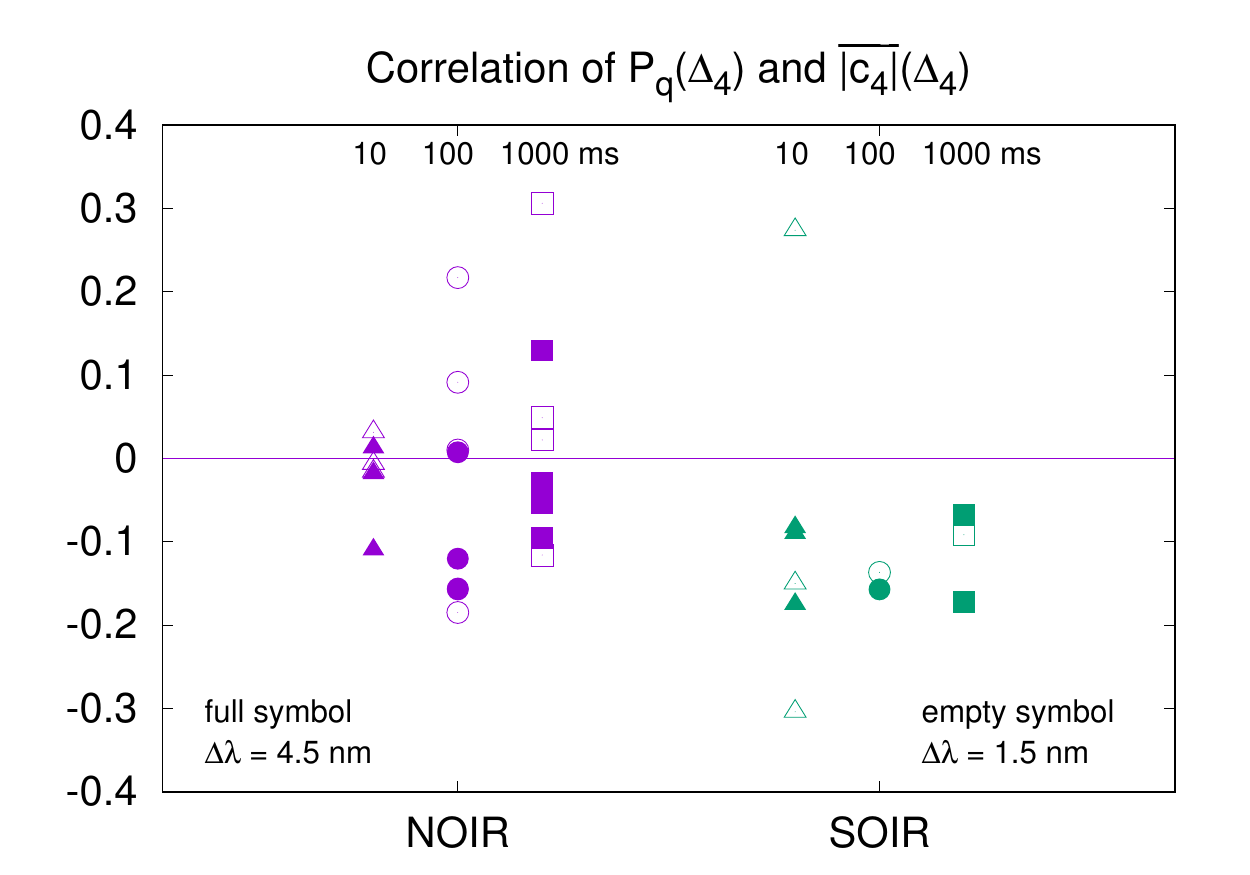}
 \caption{
 Pearson correlation between the average $\overline{|c_{4}(\Delta_{4})|}$
of the four-point correlation values and the number of quadruplets
$P_{q}(\Delta_{4})$ over which the correlations are computed. Values
obtained using data from NOIR and SOIR quadruplets are compared. Outcome
from experimental series at well different acquisition times ($10,100,1000$
ms) are reported. We repeated the analysis in two different cases
for the identification of the single peaks (see Methods): with $\Delta\lambda=4.5$
nm and one with $\Delta\lambda=1.5$ nm.  }
\label{fig:Histo_N_C2} 
\end{figure}

To be quantitative, we, further,  computed the Pearson correlation
coefficient $r$ between $\overline{|c_{4}(\Delta_{4})|}$ and $N_{q}(\Delta_{4})$
for both the SOIR and NOIR cases, as reported in Fig.~\ref{fig:Histo_N_C2}.
We stress that the NOIR correlations are typically evaluated on a much larger number of quadruplets than the SOIR ones, 
as in the former case one considers all possible quadruplets among modes in four different positions while in the latter only among modes at one fixed position.
The BG correlations are instead always evaluated with the same statistics as the SOIR ones at the same position, for an easier comparison. 
We observe that the value of $r$ is (i) always relatively small $\lesssim0.3$,
(ii) without a definite value over different experiments and (iii) of the same order of magnitude in the two cases
SOIR and NOIR. This observation reinforces the conclusions of the
previous sections, showing that our result of a dependency of $c_{4}$ from $\Delta_{4}$ cannot be explained as a small sample size effect.

\subsection{Mode-locking dependence on pumping power}

\begin{figure}[h!]
\centering 
\includegraphics[width=0.50\textwidth]{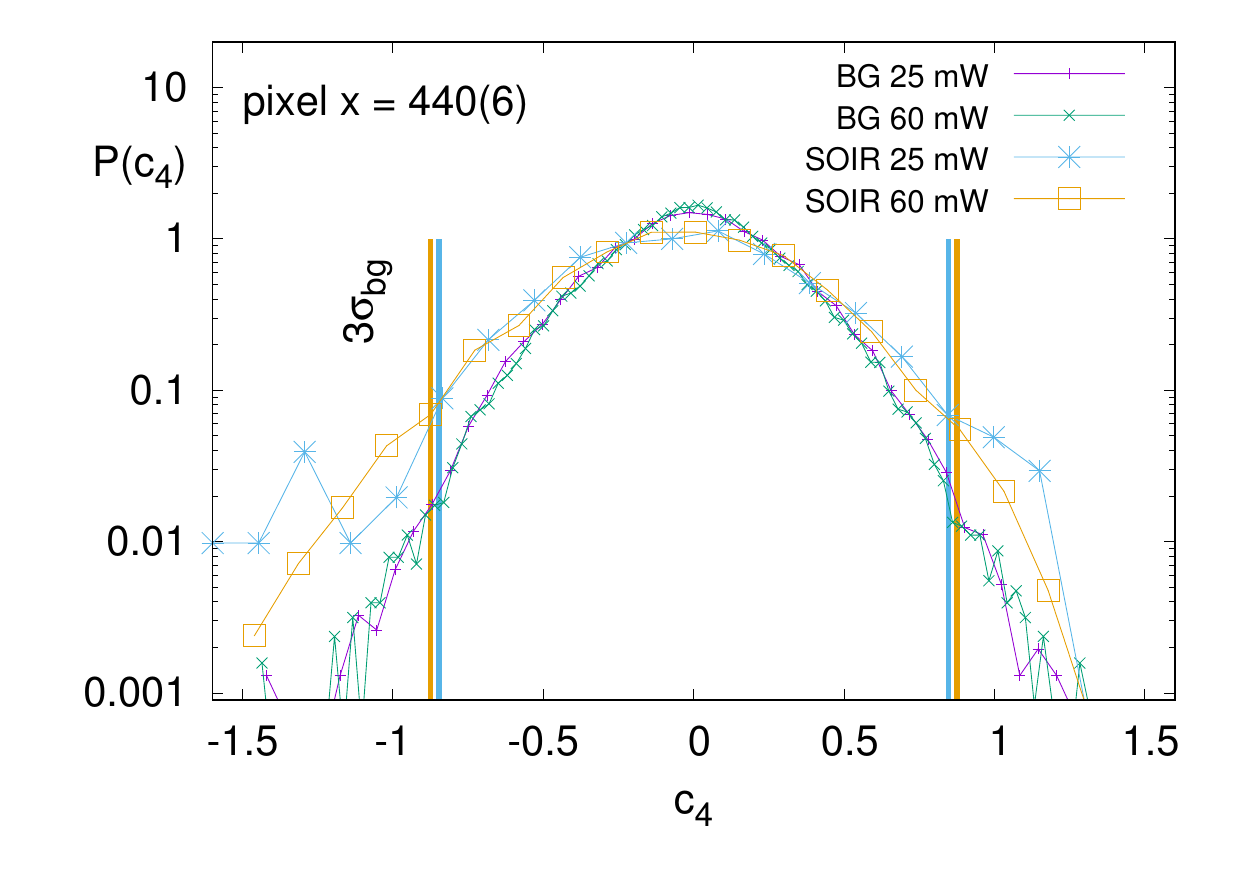}
\caption{Comparison between $P(c_{4})$ of SOIR and BG correlations at different
pumping power $P=25$ mW and $60$ mW at the same resonance at the same point 
$x=400(5)~\mu$m for acquisition time $100$ ms (1000 shots). At both
powers the SOIR distributions have wider tails than the background
ones. Both distributions do not appear to change with the pumping
power. }
\label{fig:ML_vs_P} 
\end{figure}

 Eventually we have carried out our analysis on
the same sample for different powers of the pump laser and explored
if and how non-linear mode-coupling depends on pumping. In Fig.~\ref{fig:ML_vs_P} we compare the probability distributions of the
background 4-peak correlations and the SOIR correlations at given
resonance locations at two different pumping powers, $25$ and $60$
mW. The same effect of long tails and non Gaussianity is observed
at all powers for which clear intensity resonances can be resolved
in space and wavelength and, once the mode resonances are there, no
further dependence on the external power is observed.
\label{sec:P}
\label{app:power}

\newpage
\subsection*{{References}}

\bibliographystyle{naturemag}
\bibliography{Lucabib}

\vskip 5 mm

{\bf{Ackowledgements}}

The authors thank D. Ancora, G. Gradenigo, M. Leonetti, A. Marruzzo  for useful discussions. 
The research leading to these results has received funding from the Italian Ministry of Education, University and Research under the PRIN2015 program, grant code 2015K7KK8L-005 and the European Research Council (ERC) under the European Union's Horizon 2020 research and innovation program,  project ElecOpteR Grant Agreement No. 780757 and project LoTGlasSy, Grant Agreement No. 694925. 

\vskip 5 mm

{\bf {Author contributions}}
G.L. and B.S.F. performed the measurements on the random lasers; G.L.  prepared the samples; G.L. and D.S. designed the experimental setup; D.S. coordinated the experimental work; 
F.A. and L.L. proposed the theoretical framework and performed the data analysis; L.L.. prepared the manuscript with input from F.A., D.S. and G.L. All authors contributed to the discussion of the data and to the final draft of the manuscript.
\vskip 5 mm

{\bf{Competing financial interests}}

The authors declare no competing financial interests.

\end{document}